\documentclass[twocolumn,
superscriptaddress,
prl,
showpacs,
preprintnumbers,
amsmath,amssymb]{revtex4}

\usepackage{graphicx,epsf,epsfig,ulem}% Include figure files
\usepackage{epsfig}
\usepackage{bm}% bold math
\usepackage{subfigure}
\usepackage{color}

\begin{document}

\title{High precision quantum control of single donor spins in silicon}% Force line breaks with \\

\author{Rajib Rahman}
\affiliation{Network for Computational Nanotechnology, Purdue University, West Lafayette, IN 47907, USA}

\author{Cameron J. Wellard}
\affiliation{Center for Quantum Computer Technology, School of Physics, University of Melbourne, VIC 3010, Australia}

\author{Forrest R. Bradbury}
\affiliation{Department of Electrical Engineering, Princeton University, Princeton, NJ 08544, USA}

\author{Marta Prada}
\affiliation{Network for Computational Nanotechnology, Purdue University, West Lafayette, IN 47907, USA}

\author{Jared H. Cole}
\affiliation{Center for Quantum Computer Technology, School of Physics, University of Melbourne, VIC 3010, Australia}

\author{Gerhard Klimeck}
\affiliation{Network for Computational Nanotechnology, Purdue University, West Lafayette, IN 47907, USA} 
\affiliation{Jet Propulsion Laboratory, California Institute of Technology, Pasadena, CA 91109, USA}

\author{Lloyd C. L. Hollenberg}
\affiliation{Center for Quantum Computer Technology, School of Physics, University of Melbourne, VIC 3010, Australia}

\date{\today} 

\begin{abstract} 
The Stark shift of the hyperfine coupling constant is investigated for a P donor in Si far below the ionization 
regime in the presence of interfaces using Tight-binding and Band Minima Basis approaches and 
compared to the recent precision measurements. The TB electronic structure calculations included over 3 million atoms. 
In contrast to previous effective mass based 
results, the quadratic Stark coefficient obtained from both theories agrees closely with the experiments. 
This work represents the most sensitive and precise comparison between theory and experiment for single 
donor spin control. It is also shown that there is a significant linear Stark effect for an impurity near the interface, 
whereas, far from the interface, the quadratic Stark effect dominates. %The qualitative behaviour of the linear and 
%quadratic Stark coefficients is explained by means of a perturbative model. The numerical results emphasize the 
%need for theories that go beyond the effective mass approximations to model impurities accurately. 
Such precise control of single donor spin states is required particularly in quantum computing applications of 
single donor electronics, which forms the driving motivation of this work.
\end{abstract} 

\pacs{71.55.Cn, 03.67.Lx, 85.35.Gv, 71.70.Ej}

\maketitle

Solid state quantum computer architectures have received considerable attention 
in recent years due to their promise of scalability and utilization of the vast 
knowledge and experience of the semiconductor fabrication industry. Several 
proposals employ the electronic states of phosphorus donors in silicon 
to encode qubits based on nuclear spin {\cite{Kane.nature.1998}}, 
electronic spin {\cite{Vrijen.pra.2000, Hill.prb.2005, DeSousa.pra.2004}}, or the electronic states of a 
singly ionized two-donor system {\cite{Hollenberg.prb.2004.1}}. Two-dimensional 
architectures for such qubits have been proposed to demonstrate scalability 
strategies {\cite{Hollenberg.prb.2006}}. 

In Kane's original nuclear spin qubit proposal {\cite{Kane.nature.1998}}, and the more recent
semi-global control electron spin proposal {\cite{Hill.prb.2005}}, single-qubit operations are performed by 
electrically controlling the hyperfine interaction between the donor electron 
spin and the nuclear spin of the P impurity. To implement rigorous large-scale 
quantum algorithms protected by Quantum Error Correction (QEC), the donor wave function dependent hyperfine coupling 
constant needs to be controlled to a very high degree of precision. Recent experiments have made 
tremendous inroads towards achieving such control of donor spin. In Ref {\cite{Brandt.Nature.2006}} coherent oscillations of P 
donor electron spins were measured by electrically detected magnetic resonance. Similar techniques were used in Ref {\cite{Brandt.prl.2006}} 
to invesitgate the hyperfine interaction in the presence of lattice strain for Si grown on $\textrm{Si}_{1-x}\textrm{Ge}_x$.
This work is motivated by a recent ESR experiment {\cite{Bradbury.prl.2006}} that measured the 
Stark shifts for ${}^{121} \textrm{Sb}$ donors in silicon buried at a depth of 150 nm from an interface. 
A quadratic dependence of the hyperfine coupling constant and electron g-factor on the electric field 
was reported far below the ionization regime. A weak linear Stark effect was also reported. In comparison with theory, effective mass theory (EMT)
inferred from the work of Friesen {\cite{Friesen.prl.2005}} predicts a quadratic Stark shift some ten times greater than the measured result.
This letter investigates the Stark shift in the experimentally probed regime well below the ionization threshold
and shows that two sophisticated methods,  namely Tight-Binding (TB) and Band Minima Basis (BMB), are able to predict 
the system control parameters in contrast with EMT. 
The Sb and P donors in Si can be treated as analogous because their ground state binding energies 
differ only by 2.8 meV, and their 1s manifold splittings due to valley-orbit coupling and local 
perturbations are also similar in nature {\cite{Ramdas.progphys.1981}}. This is important as both the 
TB and BMB methods employed here are optimized for P donors. This agreement between theory and experiment
for a non-trivial measure of wave function control, by two distinct theoretical approaches, represents the most
precise test of our understanding of donor systems. The current level of theoretical understanding of the system reported here, 
in conjunction with the advancement in single atom fabrication technologies {\cite{Schofield.prl.2003}} and the above-mentioned experiment, 
represents one of the first steps towards experimental realization of high precision control of donor spins in Si.

Most of the approaches based on EMT have involved using hydrogenic 
envelope functions {\cite{Kohn.physrev.1955}} with Bohr radii calculated analytically or fitted variationally 
considering the anisotropic effective masses of Si. While Kettle {\it{et al}} {\cite{Kettle.prb.2003}} used EMT to study 
the effects of electric fields and interfaces non-perturbatively with a TCAD gate potential, 
this work ignored valley-orbit (VO) interaction responsible for lifting the degeneracies of 
the 1s manifold. A number of authors {\cite{Debernardi.prb.2006, Smit.prb.2003, Friesen.prl.2005}} used EMT to investigate 
the donor energy spectrum using symmetry arguments, perturbation theory, single or the multi-valley Schr\"{o}edinger equation.
In Ref {\cite{Friesen.prl.2005}}, Friesen considered 
valley-orbit coupling and field effects in the effective mass formalism, 
to solve a multi-valley Schr\"{o}dinger equation for the six 1s states in the valley basis. 
Martins {\it{et al}} {\cite{Martins.prb.2004}} used a TB $sp^{3}s^{*}$ second nearest neighbor model to study effects of electric fields. 
In Ref {\cite{Wellard.prb.2005}} the BMB approach was proposed and applied to the same problem. 
Calder\'{o}n {\it{et al}} {\cite{Calderon.prl.2006}} used EMT to investigate shuttling times of the donor electron between a donor-bound state and a 
surface bound state. % at the interface in the ionization regime, and concluded the feasibility of establishing interface quantum control over donor spins.
Both TB and BMB approaches consider core effects and VO coupling explicitly to model impurities. The TB and the BMB methods 
are very different from each other in the sense that TB represents the Hamiltonian in real space while BMB represents it in momentum space. 
The close agreement of the results from two such different methods of calculations, as shown in this work, 
emphasizes the fundamental nature of the impurity physics. The work presented here resolves significant limitations of the previous treatments by including 
realistically large device components with over 3 million atoms in an atomistic treatment under realistic boundary conditions. 
This is also one of the first theoretical works on the Stark effect for P impurities to be
benchmarked with a relevant experiment. 
While most theoretical approaches have concentrated on donor spectrum and wave functions near the ionization regime, 
little work has been done to explore the linear and quadratic Stark shifts of the hyperfine coupling far below the ionization regime 
for realistically large systems in the presence of interfaces.

The spin Hamiltonian of a donor electron in an applied electric field $\vec \varepsilon$ and a magnetic field $B_{0}\hat{z}$ is, 
\begin{equation} \label{eq:spinham} 									
H_{z}=g(\vec \varepsilon){\mu_{B}}B_{0}S_{z} + A(\vec \varepsilon)I_{z}S_{z}
\end{equation}
where $A(\vec \varepsilon)$ and $g(\vec \varepsilon)$ are the hyperfine coupling constant and electron g-factor, respectively, 
as a function of electric field. The first term in the Hamiltonian is the spin-orbit Stark shift, 
while the second term is the hyperfine Stark shift. An electric field distorts the shape 
of the donor wave function, and modifies its angular momentum, which in turn modifies the g-factor. 
The field also pulls the donor wave function away from the impurity site, reducing $A(\vec \varepsilon)$  
which is proportional to $|\Psi(\vec \varepsilon, \vec{r_{0}})|^{2}$, where $\vec{r_{0}}$ is the impurity site. 
The change in $A(\vec \varepsilon)$ is parameterized as, 
\begin{equation} \label{eq:stark_eq}
{\Delta}A(\vec{\varepsilon}) = A(0)(\eta_{2}{\varepsilon}^{2} + \eta_{1}\varepsilon)
\end{equation}
The quadratic Stark coefficient $\eta_{2}$ was measured to be $-3.7 \times 10^{-3}$ $\mu$m$^{2}$/V$^{2}$ from the experiment by Bradbury {\it{et al}} {\cite{Bradbury.prl.2006}}.
The EMT result inferred from  {\cite{Friesen.prl.2005}} in Ref {\cite{Bradbury.prl.2006}} was $-2 \times 10^{-2}$  $\mu$m$^{2}$/V$^{2}$.
Since the spin-orbit Stark effect is small for practical magnetic fields,  this letter focuses on the behavior of 
$A(\vec \varepsilon)$ only. An impurity placed far away from interfaces in an unstrained Si lattice 
experiences a dominant quadratic Stark effect for both $A(\vec \varepsilon)$ and $g(\vec \varepsilon)$. 
However, presence of nearby interfaces or lattice strain can produce a non-negligible linear Stark 
effect {\cite{Guichar.prb.1972, Wilson.physrev.1961}}.

The origin of the linear and quadratic Stark effects can be explained qualitatively from perturbation theory.
The unperturbed ground state wave function for an impurity placed far away from interfaces has an approximate even symmetry. 
First order perturbation theory does not yield any energy correction if the perturbing Hamiltonian of a constant electric field is of odd symmetry. 
However, second order perturbation theory produces a quadratic dependence of the corrected energy on the field. 
If the impurity is very close to the interface, the even symmetry of the unperturbed wave function is broken. 
In such a case, first order perturbation theory also yields a linear dependence of the corrected energy on the electric field.

TB and BMB methods are separately used to calculate $A(\vec {\varepsilon})$.  From each calculation, the ground 
state wave function $\Psi(\vec{\varepsilon}, \vec{r})$ is calculated in the presence of the field $\vec{\varepsilon}$. 
The hyperfine coupling $A(\vec \varepsilon)$ is determined by,
\begin{equation} \label{eq:AE} 
\frac{A(\vec{\varepsilon})}{A(0)}=\frac{|\Psi(\vec{\varepsilon}, \vec {r_{0}})|^{2}}{|\Psi(0, \vec{r_{0}})|^{2}}
\end{equation}
%\textbf{
%Since the full Hamiltonian is solved non-perturbatively, the effects of the excited states, if any, are inherently included in the GS. Hence, the calculations here are 
%concerned with $A(\vec {\varepsilon})$ of GS. 
In the range of low electric fields considered here, the excited states (E and T) are not expected to affect $A(\vec {\varepsilon})$ since 
this manifold is separated by 10 meV from the ground A$_1$ state. %}
While EMT based approaches are concerned with contributions from valley-minima states only, TB and BMB approaches 
consider a more extensive Bloch structure of the material. 
%One of the main differences between EMT, BMB and TB arises from the fact that TB and BMB approaches 
%consider a more extensive %full Bloch structure of Si than EMT based approaches, which only consider contribution from the valley-minima k-states. 
TB and BMB also include valley-orbit (VO) interaction consistently, while most EMT treatments
of VO are in violation of some of the assumptions made to derive the effective mass equation {\cite{Wellard.prb.2005}}.

The BMB technique solves the Hamiltonian including an external electric field in a large but truncated basis of pure crystal Bloch 
states near the conduction band minima of a host %\textbf{
obtained by the pseudo-potential method. %}
Since the discretized Schr\"{o}dinger equation is solved in reciprocal 
space, the Fourier transform of the impurity potential is used. Near the impurity core, the potential is modified 
from its Coulombic nature to include central cell effects. 
The corrected potential produces a broader Fourier spectrum in k-space and couples different valleys 
to lift the 1s degeneracies. The core potential in momentum (q) space used in {\cite{Wellard.prb.2005}} is,
\begin{equation} \label{eq:BMB_pot}
U_{cor}(q)=\frac{1}{\pi^{2}\kappa}(\frac{{\kappa} \zeta q^{2}}{q^{2}+{\alpha}^{2}}+\frac{\kappa(1-\zeta)q^{2}}{q^{2}+{\beta}^{2}}-\frac{q^{2}}{q^{2}+{\gamma}^2})
\end{equation}
where $\kappa$ is the dielectric constant of Si, assuming a value of $11.9\varepsilon_{0}$. The form and the parameters $(\zeta,\alpha,\beta,\gamma)$ of this potential have been taken from Pantelides {\cite{Pantelides.prb.1974}}. An overall quenching factor provides a one-parameter fit to the ground state energy from which the excited state energies and degeneracies compare well with experiment. After inclusion of the external field the hyperfine coupling strength is obtained. 

The standard technique in semi-empirical TB {\cite{Slater.physrev.1954}} is to optimize a parameter set to accurately reproduce the bulk band structure of a host. Once this is done, the same set is used for atomistic modeling of any devices made of that host. 
%In the Tight-Binding technique {\cite{Slater.physrev.1954}}, 
For this work, the TB parameter set of  $sp^{3}d^{5}s^{*}$ nearest-neighbor spin model for Si was optimized by genetic 
algorithm {\cite{Boykin.prb.2004}}. %to fit the bulk band structure of Si using the $sp^{3}d^{5}s^{*}$ nearest-neighbor spin model. 
The donor potential used is, 
\begin{eqnarray} 
V(\vec{r})& =& \frac{e}{4\pi\kappa|\vec{r}-\vec{r_{0}}|}, \qquad \vec{r} \neq \vec{r_{0}}  \\ 
V(\vec{r})& = &U_{0}, \qquad \qquad \qquad \vec{r} = \vec{r_{0}}
\end{eqnarray}
The on-site core correcting potential $U_{0}$ is adjusted and found to be 4.33 eV to obtain the experimental 
ground state energy of 45.6 meV for Si:P. % \textbf{
%Although the impurity potential used reproduces the P spectrum well, 
%t may be of interest in future work to investigate the effect of additional core-corrections similar to Eqn (4).%}
The potential due to the electric field is added to the diagonal of the 
Hamiltonian. The real space Hamiltonian with closed boundary conditions is solved by a parallel Lanczos 
algorithm in NEMO 3D {\cite{Klimeck.cmes.2002}} for the donor wave function. 

%\textbf{
Both methods described here are applicable to other hosts and impurities. To describe other Group V impurities, the hydrogenic coulomb 
potential is still valid in the bulk, but the core corrections need to be adjusted to reflect correct experimental binding energies. 
%To model other host materials in BMB, the basis of Bloch wave functions used to solve the BMB Hamiltonian can be modified using standard ab initio methods. The standard technique in semi-empirical TB is to optimize a parameter set to accurately reproduce the bulk band structure of a host. Once such an optimized set is found, it is used for atomistic modeling of any devices made of that host.
%}

\begin{figure}[htbp]
\center\epsfxsize=3.4in\epsfbox{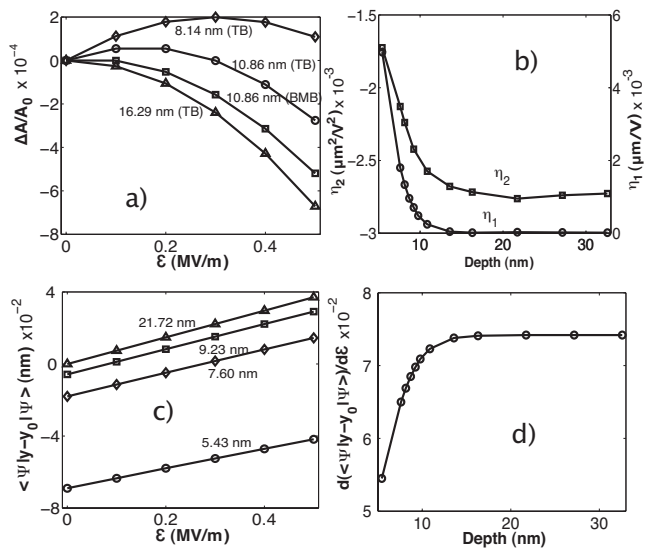}
\caption{(a) Electric field response of hyperfine coupling at various 
impurity depths (BMB and TB). (b) Quadratic (LH axis) and 
linear (RH axis) Stark coefficients with depth (TB). (c) Mean position of the ground state 
electron distribution (dipole moment) as a function of the electric field (TB). (d) The 
electric field gradient of the dipole moments (i.e. the slopes of the lines in (c))  
with respect to depth (TB). 
} \end{figure} 

Fig. 1 summarizes the effects of the electric field and the interface on the donor electron. 
The TB calculations use a domain of 32 nm $\times$ 65 nm $\times$ 32 nm Zincblende lattice with 3.45 million atoms. 
The distance between the impurity and the interface is varied parallel to the electric field.
The BMB calculation assumes a maximal depth of 10.86 nm from the interface, and employs a basis set of 7986 states. 
%\textbf{
The choice of this depth is dictated by the ease of computation as it is sufficiently deep to nullify surface effects while not 
too deep to make the problem computationally intractable. In TB, %both smaller and greater depths can be treated more conveniently, and therefore 
a range of depths from 5 nm to 32 nm have been considered. %}
For each TB data point, the typical computation times require about 7 hours on 20 CPUs {\cite{nanohub.note}}. 
Fig. 1(a) shows the variation of ${\Delta}A(\vec{\varepsilon})$ with electric field for various impurity depths. 
The data are fitted to the quadratic equation of (\ref{eq:stark_eq}) .
As the depth increases, the quadratic coefficient $\eta_{2}$ approaches a constant value, while the
linear coefficient $\eta_{1}$ becomes negligible (Fig. 1(b)). For small impurity depths, $\eta_{1}$ is 
comparable to $\eta_{2}$, which results in a shift of the peak of the parabola in Fig. 1(a) towards a non-zero 
electric field. If the linear Stark effect is negligible, an applied electric field has two effects on the ground 
state wave function: 1) a decrease in the peak amplitude of the wave function at the impurity site, 
reflected by a decrease in $A(\vec{\varepsilon})$ 
in Fig. 1(a) for higher depths, 2) a shift in the mean position of the wave function opposite 
the electric field, giving rise to a non-zero dipole moment 
as shown in Fig. 1(c). The dipole moments vary linearly with the electric field, and their slopes  
approach a constant value as the depth increases (Fig. 1(d)). The wave function plots of Fig. 2 also demonstrate these effects. 
The larger distortions of the wave function at higher electric fields explain the increasing dipole moments. 

\begin{figure}[htbp]
\epsfxsize=3.4in\epsfbox{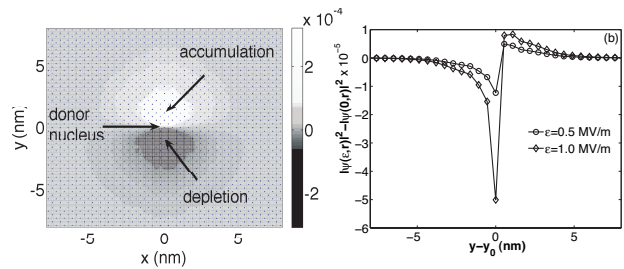}
\caption{ (a) Electric field induced differential map of donor electron wave probability density, $|\Psi(\varepsilon, z_{0})|^{2}-|\Psi(0, z_{0})|^{2}$, 
shown as a 2D cut through the impurity center at $z_{0}=16.29$ nm for $\varepsilon=0.5$ MV/m. 
The electric field is directed along the negative y axis.  (b) 1D cut though the center of the impurity parallel to the electric field showing the 
differential map of the probability density for two different electric fields.
}\end{figure}

If the impurity is close to the interface, the wave function is distorted even at zero electric field. 
This distortion comes from changes in the slope of the part of the wave function between the impurity and the interface.  
Also the mean position of the wavefunction shifts farther as the distance between the impurity and the interface is reduced. 
These effects give rise to an asymmetric charge distribution about the impurity, and a non-zero dipole 
moment is observed even at $\varepsilon=0$. This is verified by the shift of the y-intercept of the lines in Fig. 1(c) 
for small impurity depths. 
In effect, the interface behaves like an electric field pushing the donor electron away from it. The strength of this field increases 
as the impurity is placed closer to the interface. This means that a higher electric field directed 
away from the interface is needed to counteract these interface effects and to restore the decreasing behavior of $A(\vec{\varepsilon})$, 
as demonstrated in Fig. 1(a).  

Some of the numerical results can be explained qualitatively with the perturbation model using a basis 
of unperturbed impurity wave functions $ \lbrace \Psi^{0}_{0},\Psi^{0}_{1},....,\Psi^{0}_{m},...\rbrace$, where the superscript represents 
the order of correction and subscript the wave function index in order of decreasing binding energy 
(0 being the ground state). For an electric field Hamiltoninan given by $H=q \varepsilon y$, the wave function corrected to first order is expressed as,
\begin{equation} \label{eq:wave_correction}
|\Psi_{0}^{1}\rangle=|\Psi_{0}^{0}\rangle-q{\varepsilon}\sum_{m \neq 0} \sum_{i} \frac{\langle\Psi_{m,i}^{0}|y|\Psi_{0}^{0}\rangle}{E_{0}^{0}-E_{m,i}^{0}}|\Psi_{m,i}^{0}\rangle
\end{equation}
where $i$ is the degeneracy index for a state $m$. The dipole moment $D = q\langle\Psi_{0}^{1}|y|\Psi_{0}^{1}\rangle$ is then given by,
\begin{eqnarray} \label{eq:dipole_expression}
D&=&q\langle\Psi_{0}^{0}|y|\Psi_{0}^{0}\rangle
-2q^{2} \varepsilon \sum_{m \neq 0} \sum_{i} \frac{|\langle\Psi_{m,i}^{0}|y|\Psi_{0}^{0}\rangle|^{2}}{E_{0}^{0}-E_{m,i}^{0}} \nonumber \\
&&{}+q^{2} {\varepsilon}^{2} \sum_{m_{1,2} \neq 0} \sum_{i,j}  \frac{\langle\Psi_{0}^{0}|y|\Psi_{m_{1},i}^{0}\rangle\langle\Psi_{m_{2},j}^{0}|y|\Psi_{0}^{0}\rangle}{(E_{0}^{0}-E_{m_{1},i}^{0})(E_{0}^{0}-E_{m_{2},j}^{0})}   
\end{eqnarray}
The quadratic term in $\varepsilon$ can be neglected compared to the other two terms for small $\varepsilon$. This results in a linear function 
with intercept and slope depending on $\langle\Psi_{0}^{0}|y|\Psi_{0}^{0}\rangle$ and $|\langle\Psi_{0}^{0}|y|\Psi_{m,i}^{0}\rangle|^{2}{(E_{0}^{0}-E_{m,i}^{0})}^{-1}$ respectively. For an impurity far away from the interface, 
the unperturbed ground state wave function is of even symmetry, and the expression for the intercept evaluates to zero. 
Since the unperturbed wave function symmetries and eigen-energies are not affected by depth any more, the expression 
for the slope assumes a constant value. On the other hand, for small impurity depths, the unperturbed wave functions 
get distorted depending on the depth. Since the ground state has no longer an even symmetry, the intercept evaluates 
to a non-zero quantity, in consistence with Fig. 1(c). The transition probabilities between the ground state and the 
excited states are also changed due to these distortions, causing the slope of the dipole moment to vary with 
depth, in consistence with Fig. 1(d). The same reasoning explains why the quadratic Stark coefficient approaches a constant value and the linear Stark 
coefficient becomes negligible with increasing depth.    

\begin{table}
\caption{\label{tab:table1}Comparison of the quadratic Stark coefficients from experiment, EMT, BMB and TB.}
%The EMT result estimated from {\cite{Friesen.prl.2005}} is off by an order of magnitude from the experiment.}
\begin{ruledtabular}
\begin{tabular}{lll}
\hline
\textbf{Method} & \textbf{Depth (nm)} & \boldmath{$\eta_{2}$($\mu$\textbf{m}$^{2}$\textbf{/V}$^{2}$\textbf{)}} \\
\hline
Experiment (Sb) {\cite{Bradbury.prl.2006}} & 150 & $-3.7 \times 10^{-3}$ \\
\hline
EMT (P) {\cite{Friesen.prl.2005}} & $\infty$ & $-2 \times 10^{-2}$ \\
\hline
BMB (P) & 10.86 & $-2.74 \times 10^{-3}$ \\
\hline
TB (P) & 10.86 & $-2.57 \times 10^{-3}$ \\
{} & 21.72 & $-2.76 \times 10^{-3}$\\
\hline
\end{tabular}
\end{ruledtabular}
\end{table}

Table 1 gives the comparison between theory and experiment and shows that the
EMT estimate of $\eta_{2}$ differs by an order of magnitude, while TB and BMB estimates 
agree closely with each other and the experiment. The BMB estimate for a depth of 10.86 nm is already close to 
the converged TB estimate, although the peak ${\Delta}A(\vec{\varepsilon})$ at a non-zero $\varepsilon$ in Fig. 1(a) 
indicates that the BMB estimate would improve even more at greater depths. 

In summary, the Stark shift of the hyperfine coupling for Si:P in the presence of interfaces is analyzed from 
two very different theories (BMB and TB). The results of both methods are consistent and agree well with the experimental data for Si:Sb, 
thereby providing the most sensitive test of our understanding of shallow donor quantum control to date.  
%It is shown that the impurity depth can be used to engineer wave functions to control the hyperfine coupling. Since there 
%has not been any experimental work yet to study the effect of the interface on the donor, this work can also serve as a 
%general outline for designing relevant experiments. The most precise comparison made in this letter between 
%theory and experiment for single donor spin control highlights the need for approaches that go 
%beyond the effective mass approximations to model impurities accurately. 
For future work, both TB and BMB can be optimized for Sb donors, 
and the Stark shifts can be investigated in the presence of interfaces as well as lattice strain and a gate structure more closely resembling that 
of the experiment. In conclusion, we emphasize the need for very close interaction 
between theory and experiment to accomplish high precision control required in quantum computing.

\begin{acknowledgments}
Acknowledgement: This work was supported by the Australian Research Council, the Australian Government, and the US National Security Agency (NSA), Advanced Research and Development Activity (ARDA), and the Army Research Office (ARO) under contract number W911NF-04-1-0290. 
Part of this work was done at JPL, Caltech under a contract with NASA. 
NCN/nanohub.org computational resources were used.
\end{acknowledgments}
Electronic address: rrahman@purdue.edu


\begin{thebibliography}{100}
\bibitem{nanohub.note} nanoHUB.org computational resource of a
256-node 3.3GHz Pentium Irvindale PC cluster was used.
\end{thebibliography}
\end{document}